\documentstyle[here,epsfig]{mn2e}
\def\simlt{\mathrel{\rlap{\lower 3pt\hbox{$\sim$}}\raise 2.0pt\hbox{$<$}}}
\def\simgt{\mathrel{\rlap{\lower 3pt\hbox{$\sim$}} \raise 2.0pt\hbox{$>$}}}

\def\gtsima{$\; \buildrel > \over \sim \;$}
\def\ltsima{$\; \buildrel < \over \sim \;$}
\def\gtrsim{\lower.5ex\hbox{\gtsima}}
\def\lesssim{\lower.5ex\hbox{\ltsima}}
\def\url#1{{\ttfamily\def\/{/\discretionary{}{}{}}#1}}
\def\etal{{\it et al.~}}
\def\ie{{\frenchspacing i.e. }}
\def\eg{{\frenchspacing e.g. }}

\begin{document}

\newcommand{\q}{\begin{equation}}
\newcommand{\qa}{\begin{eqnarray}}
\newcommand{\qs}{\begin{eqnarray*}}
\newcommand{\nq}{\end{equation}}
\newcommand{\nqa}{\end{eqnarray}}
\newcommand{\nqs}{\end{eqnarray*}}
\newcommand{\ud}{\mathrm{d}}

\title[IGM heating by dark matter] 
{Intergalactic medium heating by  dark matter}
\author[E. Ripamonti, M. Mapelli, A. Ferrara]
{E. Ripamonti$^{1}$, M. Mapelli$^{2}$,  A. Ferrara$^{2}$\\
$^{1}$Kapteyn Astronomical Institute, University of Groningen, Postbus
800, 9700 AV, Groningen, The Netherlands; {\tt ripa@astro.rug.nl}\\
$^{2}$SISSA, International School for Advanced Studies, Via Beirut 4,
34100, Trieste, Italy\\}

\maketitle \vspace {7cm }

\begin{abstract}
We derive the evolution of the energy deposition in the intergalactic
medium (IGM) by dark matter (DM) decays/annihilations for both sterile
neutrinos and light dark matter (LDM) particles.  At $z > 200$ sterile
neutrinos transfer a fraction $f_{\rm abs}\sim{} 0.5$ of their rest mass
energy into the IGM; at lower redshifts this fraction becomes
$\lesssim{} 0.3$ depending on the particle mass. The LDM particles can
decay or annihilate. In both cases $f_{\rm abs}\sim{} 0.4-0.9$ at high
($> 300$) redshift, dropping to $\approx 0.1$ below $z=100$. These
results indicate that the impact of DM decays/annihilations on the IGM
thermal and ionization history is less important than previously
thought.  We find that sterile neutrinos (LDM) decays are able to
increase the IGM temperature by $z=5$ at most up to $4$~K ($100$~K),
about 50-200 times less than predicted by estimates based on the
assumption of complete energy transfer to the gas.
\end{abstract}

\section{Introduction}\label{introduction section}
%Part I: literature\\
According to 3-yr WMAP ({\it Wilkinson Microwave Anisotropy Probe})
results (Spergel \etal 2006), the dark matter (DM) constitutes about
20\% of the cosmic energy density. However, the nature of such elusive
component remains unclear.

As proposed DM candidates might induce drastically different
evolutionary effects depending on their properties (\eg
velocity dispersion), one hopes to be able to select suitable DM
candidates from the comparison between their predicted effects and
observations.

Potentially important cosmological effects might be induced if DM
particles either decay or annihilate, as predicted by fundamental
physics theories.  For example, sufficiently light DM particles (mass
$\simlt{}100$ MeV; Boehm \etal 2004; Ascasibar \etal 2006) can
annihilate, or decay into lighter particles (Hooper \& Wang 2004;
Picciotto \& Pospelov 2005; Ascasibar \etal 2006) remaining good DM
candidates. The products of DM decays/annihilations can be
photons, neutrinos, electron-positron pairs, and/or more massive
particles, depending on the mass of the progenitor.

The decay/annihilation of DM particles into $e^+-e^-$ pairs has been
recently invoked to explain the observation, by the SPI spectrometer
aboard ESA's INTEGRAL satellite, of an excess in the 511-keV line
emission from the Galactic bulge (Kn\"odlseder \etal 2005). Although
exotic, this idea has triggered many theoretical studies (Boehm \etal
2004; Hooper \& Wang 2004; Picciotto \& Pospelov 2005; Kawasaki \&
Yanagida 2005; Kasuya \& Takahashi 2005; Cass\'e \& Fayet 2006;
Ascasibar \etal 2006), aimed at constraining DM properties through
SPI/INTEGRAL observations.

The products of decays and/or annihilations are expected to interact
with the intergalactic medium (IGM), transferring part of their energy.
% being partially absorbed by it.
If so, DM decays and annihilations might change the IGM thermal/ionization
history in a sensible and detectable way. Various flavors of this
mechanism have been investigated in a considerable number of studies
(Hansen \& Haiman 2004; Chen \& Kamionkowski 2004; Pierpaoli 2004;
Padmanabhan \& Finkbeiner 2005; Mapelli \& Ferrara 2005; Biermann \&
Kusenko 2006; Mapelli, Ferrara \& Pierpaoli 2006, hereafter MFP06; Zhang
\etal 2006).

The above studies, though, made the simplifying assumptions that either
(i) the energy injected by DM decays/annihilation is {\it entirely}
absorbed by the IGM (Hansen \& Haiman 2004; Pierpaoli 2004; Biermann \&
Kusenko 2006; MFP06), or (ii) leave the absorption efficiency as a free
parameter (Padmanabhan \& Finkbeiner 2005; Zhang \etal 2006), or (iii)
make a partial treatment of the physical processes responsible for the
energy redistribution (Chen \& Kamionkowski 2004; Mapelli \& Ferrara
2005).

In this paper we model in detail, for the first time, the physical
processes governing the interaction between the IGM and the
decay/annihilation products, and we derive the
fraction of energy actually absorbed
% effective energy absorbed fraction
(Section \ref{section method}). We
restrict our analysis to the case in which the decay/annihilation
products are photons, electron-positron pairs, or neutrinos (which are
assumed to have negligible interactions with matter), because of the
uncertainties in modeling the cascade associated with more massive
product particles. For photons (Section \ref{photon subsection}) we
include the effects of Compton scattering and photo-ionization; for
pairs, the relevant processes are inverse Compton scattering,
collisional ionizations, and positron annihilations (Section \ref{pair
subsection}).

Our model exhaustively describes the behavior of relatively light DM
candidates (i.e. warm and cold DM particles with mass lower than
$\sim{}100$ MeV), whose only decay/annihilation products are photons,
pairs and neutrinos. As an application, we explore the effects of
sterile neutrinos (mass 2-50 keV) and of viable light dark matter
particles (LDM; mass 1-10 MeV) on the IGM thermal and ionization history
(Section \ref{section applications}).

We adopt the best-fit cosmological parameters after the 3-yr WMAP
results (Spergel \etal 2006), i.e. $\Omega{}_{\rm b}=0.042$,
$\Omega{}_{\rm M}=0.24$, $\Omega{}_{\rm DM}\equiv{}\Omega{}_{\rm
M}-\Omega{}_{\rm b}=0.198$, $\Omega{}_\Lambda{}=0.76$, $h=0.73$,
$H_0=100\,{}h$ km s$^{-1}$ Mpc$^{-1}$.

%\section{METHOD}
\section{Energy injection in the IGM}\label{section method}
We consider the case of DM decays and annihilations resulting in the
direct production of photons, electron-positron pairs, and/or
neutrinos, and study the absorption of the energy of such particle products 
by the IGM.

Some fraction of the energy of the newly created particles is immediately
absorbed; the remainder will remain in the form of a {\it background}, 
eventually absorbed at later times.

In practice, at each redshift $z$, it is convenient to distinguish the
particles which were produced by DM decays and/or annihilations
according to their ``production redshift'' $z'$ and energy $E$. We
define $n(z,z')$ as the number (per baryon) of  product
particles which at redshift $z$ can still inject energy in the IGM, and
were produced at redshift $z'\geq z$. Their energy spectrum is
${{dn}\over{dE}}(z,z',E)$.

The rate of energy absorption per baryon in the IGM at redshift $z$ is
the sum of the contributions of all ``product particles'',
obtained by the integration over production redshift and energy
\begin{equation}
\epsilon(z) = \int_z^{z_{\rm max}}{dz'\,
  \int{dE\, {{dn}\over{dE}}(z,z',E)\, E\, \phi(z,E)}}
\label{energy injection iper-general}
\end{equation}
where $\phi(z,E)$ is the fraction of the energy $E$ of a particle which
is absorbed by the IGM per unit time, calculated at redshift $z$.
We assume $z_{\rm max}$=1100, which is approximately the redshift of
the last scattering surface (Spergel \etal 2006). 

However, using equation (\ref{energy injection iper-general}) is both
complicated and unnecessary: in fact, we know that the energy spectrum
of ``fresh'' DM decay/annihilation products is essentially
mono-energetic, and it is reasonable to expect that it will remain
peaked at the average energy. Therefore, we assume that the energy
integration inside equation  (\ref{energy injection iper-general}) can be
safely eliminated by using the average energy $\bar{E}(z,z')$ of the particles
\begin{equation}
\bar{E}(z,z') = {1\over{n(z,z')}} \int{dE\, {{dn}\over{dE}}(z,z',E)\, E}
\end{equation}
so that equation (\ref{energy injection iper-general}) becomes
\begin{equation}
\epsilon(z) =\int_z^{z_{\rm max}}{dz'\, n(z,z')\, \bar{E}(z,z')\,
\phi(z,\bar{E}(z,z'))}.
\label{energy injection general}
\end{equation}
The evolution of $n(z,z')$, $\bar{E}(z,z')$, and $\phi(z,E)$ obviously
depends on the type of DM decay/annihilation product we are
considering. If the product particles are neutrinos $\phi_\nu(z,E)=0$;
the cases of photons and electron-positron pairs are discussed in the
following subsections.

\subsection{Photons}\label{photon subsection}
If the DM decays or annihilations result in the production of photons,
the only two important energy loss mechanisms in the considered energy
and redshift ranges ($25\,{\rm eV}\lesssim E \lesssim 50\,{\rm MeV}$;
$z\lesssim 1000$ ) are ionizations and Compton scattering on
cold matter (see Zdziarski \& Svensson 1989; hereafter ZS89). All the
energy lost by the photons is absorbed by the IGM, so we can write
\begin{equation}
\phi_\gamma(z,E)=\phi_{\gamma,{\rm ion}}(z,E) +
\phi_{\gamma,{\rm com}}(z,E).
\end{equation}

The photo-ionization term can be expressed as
\begin{equation}
\phi_{\gamma,{\rm ion}}(z,E) = {{\sigma_{\rm He+H}(E)}\over{16}} \,
 N_{\rm b}(z)\, c,
\label{gamma ionization losses}
\end{equation}
where $N_{\rm b}(z) \simeq 2.5\times 10^{-7} (1+z)^3 (\Omega_{\rm b}
h^2/0.0224)\;{\rm cm^{-3}}$ is the number density of the
baryons  at redshift $z$, and
\begin{equation}
\sigma_{\rm He+H}(E) = \sigma_{\rm He} + 12\sigma_{\rm H}
\simeq 5.1\times10^{-20}
\left({E\over{\rm 250\,eV}}\right)^{-p}{\rm cm^2}
\end{equation}
(with $p=3.30$ for $E>250\,{\rm eV}$, $p=2.65$ for $25\,{\rm eV}\leq
E\leq 250\,{\rm eV}$) is the photoionization absorption cross-section
per helium atom of the cosmological mixture of H and He (see equation
3.2 of ZS89).  Equation (\ref{gamma ionization losses}) implicitly
assumes that the IGM is mostly neutral, which is true between the
hydrogen recombination at $z\simeq 1100$, and its reionization at
$z\sim6-15$.
 
The Compton scattering term is
\begin{equation}
\phi_{\gamma,{\rm com}}(z,E) = [\sigma_{\rm T}\, \xi\, g(\xi)] \,
N_{\rm e}(z) \, c
\label{gamma compton losses}
\end{equation}
where $\xi \equiv E/(m_e c^2)$, $N_{\rm e}(z)$ is the electron
number density, and the product $\sigma_{\rm T}\, \xi\, g(\xi)$ gives
the average fraction of energy which is lost by a photon for each
electron on its path ($\sigma_{\rm T}\simeq 6.65\times10^{-25}\,{\rm
cm^{-2}}$ is the Thomson cross-section; see equation 4.9 of ZS89 for the
definition of the function $g(\xi)$).  In equation (\ref{gamma compton
losses}) the electron number density accounts for both free and bound
electrons, as the energy losses due to Compton scattering only become
important when $E$ is so high that the interaction is insensitive to
whether an electron is bound or free (Chen \& Kamionkowski 2004). Then,
$N_{\rm e}(z) = [(1+2f_{\rm He})/(1+4f_{\rm He})]\, N_{\rm
b}(z)\simeq 0.88\, N_{\rm b}(z)$, where $f_{\rm He}=0.0789$ is the
helium-to-hydrogen number ratio.

%In
%equation \ref{gamma compton losses} the baryon density appears 
%instead of the electron density, as the energy losses due to Compton
%scattering only become important when $E$ is so high that the
%interaction is insensitive to whether an electron is bound or free (Chen
%\& Kamionkowski 2004).

We assume that the energy transfer through ionization results in a
photon loss, whereas Compton scatterings reduce the average energy of
the photons without changing their number. So, for $z<z'$ the equations
for the cosmological evolution of $n(z,z')$ and $E(z,z')$ are
\begin{eqnarray}
{{n(z+dz,z')}\over{n(z,z')}} & = &
1-\phi_{\gamma,{\rm ion}}(z,\bar{E}(z,z')){{dt}\over{dz}} dz \\
{{\bar{E}(z+dz,z')}\over{\bar{E}(z,z')}} & = & 
1- \phi_{\gamma,{\rm com}}(z,\bar{E}(z,z'))
    {{dt}\over{dz}} dz + {{dz}\over{1+z}}
\end{eqnarray}
where the energy equation also keeps into account the cosmological
redshifting of photons (also note that $dt/dz$ is negative).

These equations need to be supplemented with the injection of new
photons at each redshift $z=z'$:
\begin{eqnarray}
n(z'+dz,z') & = & \zeta_1\, \dot{n}_{\rm DM}(z')\, {{dt}\over{dz}} dz
\label{n injection photons}\\
\bar{E}(z'+dz,z') & = &
\zeta_2\, m_{\rm DM}\, c^2\, \left({1+{{dz}\over{1+z}}}\right)
\label{e injection photons}
\end{eqnarray}
where $\dot{n}_{\rm DM} (z')$ is the rate of decrease of the number of
DM particles per baryon at redshift $z'$, and $m_{\rm DM}$ is the mass of a
DM particle (see Section \ref{fabs subsection}). $\zeta{}_1$ and $\zeta{}_2$
are numerical coefficients which depend on the considered DM particle
and on the details of its decay or annihilation.

\subsection{Pair production}\label{pair subsection}
Even if the loss of kinetic energy from electrons and positrons can be
treated exactly in the same way, the annihilation probability is
negligible for the electrons, but must be kept into account for the
positrons. For this reason, when DM decays/annihilations result in the
production of an electron-positron pair, it is useful to distinguish between electrons and
positrons. 

In the case of
annihilations we neglect other processes, such as the internal
Bremsstrahlung, which affect $\epsilon(z)$ only in a minor way (see
Appendix A).

Note that in the following we will always include the rest
energy $m_e c^2$ as part of the energy $E$ of the particle.

\subsubsection{Electrons}
Electrons can transfer their kinetic energy to the IGM through
collisional ionizations and ionizations by inverse Compton up-scattered
cosmic microwave background (CMB) photons. Here we neglect the energy
loss through synchrotron radiation, because the inverse Compton
mechanism is more efficient by a factor $U_{\rm CMB}/U_{\rm B}$ (where
$U_{\rm CMB}$ and $U_{\rm B}$ are the energy densities of the CMB and of
the magnetic field, respectively), which is $\gg 1$ unless
unrealistically strong magnetic fields ($B\gtrsim10^{-5}\;{\rm gauss}$)
are assumed to exist in the IGM at $z\geq 5$.

It is necessary to remark that photons produced by the
inverse Compton mechanism are not necessarily absorbed by the IGM, and
we must distinguish between the fractional energy loss rate by electrons
\begin{equation}
\Phi_{\rm e-}(z,E)=\Phi_{\rm e,ion}(z,E) + \Phi_{\rm e,com}(z,E),
\end{equation}
and the fractional energy loss rate actually absorbed by the IGM,
\begin{equation}
\phi_{\rm e-}(z,E)=\phi_{\rm e,ion}(z,E) + \phi_{\rm e,com}(z,E)
\end{equation}

The ionization losses are completely absorbed by the IGM, so that
\begin{equation}
\begin{array}{l}
\phi_{\rm e,ion}(z,E) = \Phi_{\rm e,ion}(z,E) \simeq
{v\over E} {{2\pi e^4}\over{m_e v^2}}\,\times \nonumber\\
\{ Z_{\rm H}\, N_{\rm H}(z)\, \, 
\left[{
\ln\left({{m_e v^2 \gamma^2 T_{\rm max,H}}\over{2E_{\rm th,H}^2}}\right)
+{\mathcal D}(\gamma{})}\right]+\nonumber{}\\
%+ {1\over{\gamma^2}} -
%\left({{2\over\gamma} - {1\over{\gamma^2}}}\right)\ln{2} +
%{1\over8}\left({1-{1\over\gamma}}\right)^2}\right] + \nonumber\\
+\, Z_{\rm He}\, N_{\rm He}(z)\,
\left[{
\ln\left({{m_e v^2 \gamma^2 T_{\rm max,He}}\over{2E_{\rm th,He}^2}}\right)
%+ {1\over{\gamma^2}} -
%\left({{2\over\gamma} - {1\over{\gamma^2}}}\right)\ln{2} +
%{1\over8}\left({1-{1\over\gamma}}\right)^2
+{\mathcal D}(\gamma{})}\right]\}
\end{array}
\end{equation}
where $v$ and $e$ are the velocity and charge of the electron,
$\gamma=E/(m_e c^2)$ is the electron Lorentz factor, $E_{\rm
th,H}=13.59\;{\rm eV}$ ($E_{\rm th,He}=24.6\;{\rm eV}$) is the hydrogen
(helium) ionization threshold, $Z_{\rm H}=1$ ($Z_{\rm He}=2$) is the
hydrogen (helium) atomic number, $N_{\rm H}(z) = N_{\rm b}(z)/(1+4f_{\rm
He})$ is the H number density, $N_{\rm He}(z) = N_{\rm b}(z)\, f_{\rm
He}/(1+4f_{\rm He})$ is the He number density, 
${\mathcal D}(\gamma{})={1\over{\gamma^2}} -
\left({{2\over\gamma} - {1\over{\gamma^2}}}\right)\ln{2} +
{1\over8}\left({1-{1\over\gamma}}\right)^2$
, and
\begin{equation}
\begin{array}{l}
T_{\rm max,H}={{2\gamma^2 m_{\rm H}^2 m_e v^2}
\over{m_e^2 + m_{\rm H}^2+2\gamma m_e m_{\rm H}}} \nonumber \\
T_{\rm max,He}={{2\gamma^2 (4 m_{\rm H})^2 m_e v^2}
\over{m_e^2 + (4 m_{\rm H})^2+2\gamma m_e (4 m_{\rm H})}}
\end{array}
\end{equation}
(cfr. Longair 1992, and also Lang 1999, and Chen  \& Kamionkowski 2004).

The total Compton fractional losses are given by
\begin{equation} 
\Phi_{\rm e,com}(z,E) = \left({1\over E}\right)
{4\over3} \sigma_{\rm T}\, c\, a_{\rm rad}\, T_{\rm CMB}(z)^4\, (\gamma^2-1),
\end{equation}
where $a_{\rm rad}\simeq 7.56\times10^{-15}\;{\rm erg\,cm^{-3}\,K^{-4}}$
is the Stefan-Boltzmann constant for the radiation energy density, and
$T_{\rm CMB}(z) \simeq 2.726\,(1+z)\;{\rm K}$ is the temperature of the
CMB radiation at redshift $z$ (cfr. Rybicki \& Lightman 1979; Longair 1992); the fractional 
loss rate that is actually absorbed by the IGM is
\begin{equation}
\phi_{\rm e,com}(z,E) = \Phi_{\rm e,com}(z,E) {{\gamma^2}\over{\gamma^2-1}}
\int_{\nu(\gamma)}^\infty {d\nu {{4\pi B_\nu(T_{\rm CMB}(z))}
\over {a_{\rm rad}\, c\, T_{\rm CMB}(z)^4}}}
\label{electron compton loss abs}
\end{equation}
where $B_\nu(T)$ is the Planck function for black-body radiation at
temperature $T$, and
\begin{equation}
\nu(\gamma)={{3\,E_{\rm th,H}}\over{4\,h_{\rm P}\,\gamma^2}}
\end{equation}
where $h_{\rm P}$ is the Planck constant.
The correction to $\Phi_{\rm e,com}$ introduced in equation
(\ref{electron compton loss abs}) amounts to neglecting photons with
pre-interaction frequencies below $\nu(\gamma)$. This is necessary
because on average the post-interaction energies of such photons are
below $E_{\rm ion,H}$, so that they are hardly absorbed by the
IGM\footnote{Lyman $\alpha$ opacity generally results only in the
scattering of the photon, rather than in its absorption, and only a
small fraction of the energy is absorbed (Furlanetto \& Pritchard 2006). %However, our
%results do not change very much if we move the $\nu(\gamma)$ threshold
%to the energy of the H Lyman $\alpha$ line.
}. 

As in the case of photons, we assume that ionization energy losses lead
to the disappearance of electrons, whereas Compton losses reduce the
average energy of the surviving electrons. Therefore, the equations
describing the evolution of the energetic electrons coming from decays and
annihilations are quite similar to those we used
for photons. For redshifts $z<z'$ we have
\begin{eqnarray}
{{n_{\rm e-}(z+dz,z')}\over{n_{\rm e-}(z,z')}} & = &
1 - \Phi_{\rm e,ion}(z,\bar{E}_{\rm e-}(z,z')){{dt}\over{dz}} dz \\
{{\bar{E}_{\rm e-}(z+dz,z')}\over{\bar{E}_{\rm e-}(z,z')}} & = & 
1 - \Phi_{\rm e,com}(z,\bar{E}_{\rm e-}(z,z'))
    {{dt}\over{dz}} dz,
\end{eqnarray}
whereas the injection of new electrons is described by
\begin{eqnarray}
n_{\rm e-}(z'+dz,z') & = & \zeta_1\, \dot{n}_{\rm DM}(z')\,
{{dt}\over{dz}} dz
\label{n injection electrons}\\
\bar{E}_{\rm e-}(z'+dz,z') & = & \zeta_2\, m_{\rm DM}\, c^2.
\label{e injection electrons}
\end{eqnarray}
where, again, $\zeta_1$ and $\zeta_2$ depend on the details of the
considered decaying or annihilating particle.

\subsubsection{Positrons}
In addition to the energy loss mechanisms of electrons, decay produced
positrons can also annihilate with thermal electrons in the surrounding
gas. The fractional energy loss due to the annihilations of positrons
of energy $E$ is
\begin{equation}
\Phi_{\rm ann}(E) = v\, N_{\rm e}(z)\, \sigma_{\rm ann}(E)
\label{annihilation total loss}
\end{equation}
where the annihilation cross-section $\sigma_{\rm ann}$ is (cfr. Beacom
\& Y\"uksel 2006)
\begin{eqnarray}
\sigma_{\rm ann}(E) & = & {{3\sigma_{\rm T}}\over{8(\gamma+1)}} \times\\
&& \left[{{{\gamma^2+4\gamma+1}\over{\gamma^2-1}}\ln(\gamma+\sqrt{\gamma^2-1})
-{{\gamma+3}\over{\sqrt{\gamma^2-1}}}}\right]. \nonumber
\label{annihilation cross section}
\end{eqnarray}

Every annihilation emits two photons, each of energy $E_\gamma \sim
{1\over2}(E+m_e c^2)$, as it involves a positron of energy $E$ and an
electron whose energy is likely to be close to $m_e c^2$. Such photons
are absorbed only if the optical depth they encounter is sufficiently
high. Here we do not follow their radiative transfer in detail, and
simply assume that the fractional energy loss which actually goes in the
IGM is
\begin{equation}
\phi_{\rm ann}(z,E) =
\Phi_{\rm ann}(z,E)\;f_1\, [1-e^{-\tau_\gamma(z,E_\gamma)}]
\end{equation}
with
\begin{equation}
\tau_\gamma(z,E_\gamma) = {{f_2}\over{H(z)}}\,
%\left[{ \sigma_T\,
%\xi_\gamma g\left(\xi_\gamma\right)\,n_{\rm b}(z)\, c}\right]
\phi_{\rm \gamma,com}(z,E_\gamma)
\label{tau e+}
\end{equation}
where $H(z)$ is the expansion rate of the universe at redshift $z$, and
$\phi_{\rm \gamma,com}(z,E_\gamma)$ is the Compton fractional energy
loss of a photon of energy $E_\gamma$, as defined in equation
(\ref{gamma compton losses}).
%the term in square brackets correspond to the Compton fractional energy
%loss of a photon of energy $E_\gamma$; see equation (\ref{gamma compton
%losses}).
Equation~(\ref{tau e+}) represents a fraction $f_2$ of the optical depth encountered by a
photon of energy $E_\gamma$ emitted at redshift $z$ and traveling an
Hubble radius, assuming that the baryonic density does not vary. The
parameters $f_1=0.91$ and $f_2=0.6$ have been chosen in order to
maximize the agreement between a full radiative transfer treatment and
our simple approximation.

Equation (\ref{annihilation cross section}) implies that annihilation
energy loss is particularly efficient for positrons with very low
kinetic energy, as $\sigma_{\rm ann}(E)\propto(\gamma-1)^{-1/2}$ when $E
\approx m_e\, c^2$ (and $\gamma$ tends to $1$).

For this reason, it is convenient to separate positrons in two different
groups: ``fast'' positrons (which effectively lose energy
through ionizations, inverse Compton, and annihilations), and
``thermal'' positrons (whose kinetic energy is so low that
annihilations are their only energy loss mechanism).

The treatment of fast positrons is very similar to that of
electrons: their fractional energy loss rate is
\begin{equation}
\Phi_{\rm e+,f}(z,E) = 
\Phi_{\rm e,ion}(z,E) + \Phi_{\rm e,com}(z,E) + \Phi_{\rm ann}(z,E),
\end{equation}
and the fractional energy loss rate actually absorbed by the IGM is
\begin{equation}
\phi_{\rm e+,f}(z,E) =
\phi_{\rm e,ion}(z,E) + \phi_{\rm e,com}(z,E) + \phi_{\rm ann}(z,E),
\end{equation}
where
%the quantities with the $f$ subscripts
%%pedices 
%only refer to ``fast''
%positrons (\eg, $E_{\rm f}$ is the average energy of the fast
%positrons), and
the ionization and Compton terms are exactly the same as
in the case of electrons.

Their evolution is described by the equations which are simple
modifications of those given for electrons, taking annihilations into
account
\begin{eqnarray}
{{n_{\rm f}(z+dz,z')}\over{n_{\rm f}(z,z')}} & = &
1-[\Phi_{\rm e,ion}(z,\bar{E}_{\rm f}(z,z'))\nonumber{}\\ && +
  \Phi_{\rm ann}(z,\bar{E}_{\rm f}(z,z'))] {{dt}\over{dz}} dz\\
{{\bar{E}_{\rm f}(z+dz,z')}\over{\bar{E}_{\rm f}(z,z')}} & = &
1-\Phi_{\rm e,com}(z,\bar{E}_{\rm f}(z,z')) {{dt}\over{dz}} dz,
\end{eqnarray}
where the ``f'' subscripts refer to ``fast'' positrons (\eg,
$\bar{E}_{\rm f}$ is the average energy of the fast positrons).  Their
injection rate and average energy at injection are identical to those
given in equations (\ref{n injection electrons}) and (\ref{e injection
electrons}) for electrons: $n_{\rm f}(z'+dz,z')=n_{\rm e-}(z'+dz,z')$,
$\bar{E}_{\rm f}(z'+dz,z')=\bar{E}_{\rm e-}(z'+dz,z')$.

The fractional energy losses (total and absorbed by the IGM) of
thermal positrons are simply
\begin{eqnarray}
\Phi_{\rm e+,t}(z,E) = \Phi_{\rm ann}(z,E)\\
\phi_{\rm e+,t}(z,E) = \phi_{\rm ann}(z,E);
\end{eqnarray}
%where the $t$ pedices refer to properties of thermal positrons.
their evolution equations are
\begin{eqnarray}
n_{\rm t}(z+dz,z') & = &
n_{\rm t}(z,z')[1-\Phi_{\rm ann}(z,\bar{E}_{\rm t}(z))] \nonumber{}\\&&+
n_{\rm f}(z,z') \Phi_{\rm e,ion}(z,\bar{E}_{\rm f}(z,z'))\\
\bar{E}_{\rm t}(z) & = & m_e c^2 + {3\over2} k_{\rm B} T_{\rm IGM}(z)
\end{eqnarray}
where the ``t'' subscripts refer to ``thermal'' positrons, $k_{\rm B}$
is the Boltzmann constant, and $T_{\rm IGM}(z)$ is the temperature of the
IGM at redshift $z$. We take the injection rate to be simply $n_{\rm
t}(z',z')=0$, as positrons are naturally ``fast'' when they are created.

In practice, these equations assume that the only mechanism which leads
to the disappearance of a positron is its annihilation; instead,
ionization energy losses simply turn a fast positron into a thermal one.

\subsection{The absorbed energy fraction}\label{fabs subsection}
The energy injection rate per baryon resulting from the integration of
equation (\ref{energy injection general}) can be expressed in the form
of the fraction $f_{\rm abs}$ of the total energy released by the DM
which is absorbed by the IGM
\begin{equation}
f_{\rm abs}(z) = {{\epsilon(z)}\over
{\dot{n}_{\rm DM}(z)\, m_{\rm DM}}\, c^2}.
\label{absorbed fraction def}
\end{equation}
In the case of decaying DM, $\dot{n}_{DM}$ (which we defined as the
decrease rate of the number of DM particles per baryon) is given by
\begin{equation}\label{DM density decay}
\dot{n}_{\rm DM}(z) = {{n_{{\rm DM},0}}\over{\tau_{\rm DM}}}
e^{{t(0)-t(z)}\over{\tau_{\rm DM}}} \simeq {{n_{{\rm
	DM},0}}\over{\tau_{\rm DM}}},
\end{equation}
where $n_{\rm DM,0}$ is the number of DM particles per baryon at
present, $\tau_{\rm DM}$ is the lifetime of a DM particle, and $t(0)$ and
$t(z)$ are the ages of the universe at present and at redshift $z$,
respectively. The leftmost equality is valid when $\tau_{\rm DM}\gg
t(0)$, which is generally the case.

Instead, in the case of annihilating DM, $\dot{n}_{\rm DM}$ is 
\begin{equation}\label{DM density annih}
\dot{n}_{\rm DM}(z) \simeq {1\over2} \, n_{{\rm DM},0}^2\,{}
N_{\rm b}(0)\,{} \langle\sigma\,v\rangle (1+z)^3,
\end{equation}
where $\langle\sigma v\rangle$ is the thermally averaged annihilation
cross-section; the $1/2$ factor is due to two reasons: first, the DM is
split in half between particles and anti-particles, and this needs to be
accounted by introducing a correction factor $1/4$. However, this must
be multiplied by 2, as each annihilation involves two DM particles.

The definition of the absorbed fraction\footnote{We remark
that the term ``absorbed fraction'' might be slightly misleading, as it
%is theoretically possible to have $f_{\rm abs}> 1$ at least at some
%$z$. This might happen in scenarios where $\epsilon(z)$ is dominated by
is theoretically possible to have $f_{\rm abs}> 1$ in scenarios where
$\epsilon(z)$ is dominated by the absorption from particles in the
``background'', rather than from the ones which were produced
recently. This might happen, for example, when $\dot{n}_{\rm DM}$
decreases very rapidly.}, 
$f_{\rm abs}$, given in equation (\ref{absorbed
fraction def}) is such that it can be easily plugged into the equations
commonly used in studies concerned with decaying and annihilating DM
%given in most of the literature about decaying and annihilating DM 
(\eg Padmanabhan \& Finkbeiner 2005; Zhang \etal 2006).  A second
advantage of the $f_{\rm abs}$ notation is that it is independent of
uncertain quantities such as $\tau_{DM}$ and $\langle\sigma\,v\rangle$.

\section{Applications}\label{section applications}
%\subsection{The contribution to ionization fraction and heating}
%The energy absorbed fraction $f_{\rm abs}$
The rate of energy absorption per baryon, $\epsilon{}(z)$, that
was derived in the previous Section, can be used to improve our knowledge
of the effects of DM decays/annihilations on cosmic reionization and
heating. In fact, most of the previous works (Chen \& Kamionkowski 2004;
Padmanhaban \& Finkbeiner 2005; MFP06; Zhang \etal 2006) derived only
an upper limit to these effects, by assuming that all the energy emitted
during the decay/annihilation is absorbed, or considering the energy
absorbed fraction, $f_{\rm abs}(z)$ as a free (and mostly unknown)
parameter.

Now, we have an estimate of $\epsilon{}(z)$ and $f_{\rm abs}(z)$ which
accounts for all the important physics involved, and we can use it to
derive the effective influence of DM decays/annihilations on the cosmic
reionization and heating. For this purpose, we ran the public
version of the code RECFAST (Seager, Sasselov \& Scott 1999, 2000),
modified to account for the energy injection from DM decays and
annihilations. The adopted procedure is mostly the same as described in
MFP06. Here, we briefly summarize the most important assumptions of this
method, pointing out the improvements of our present treatment.

The IGM is heated, excited and ionized by the energy input due to DM
decays/annihilations. It is important to note that the fraction of the
absorbed energy going into each one of these components is quite
unrelated to how the energy was deposited in the IGM in the first
place. For example, if a keV a photon ionizes an atom, the resulting
electron will generate a cascade of collisions, and the energy of the
photon will go not only into ionizations, but also into excitations and
heating.

In order to treat this process, we assume that a fraction $(1-x)/3$
(where $x$ is the ionization fraction) of the energy absorbed by the IGM
contributes to the ionizations (Chen \& Kamionkowski 2004), and that a
fraction ${\mathcal
F}(x)=\tilde{C}\,{}\left[1-(1-x^{\tilde{a}})^{\tilde{b}}\right]$ (where
$\tilde{C}$=0.9971, $\tilde{a}$=0.2663 and $\tilde{b}$=1.3163; Shull \&
van Steenberg 1985) goes into heating. This definition of ${\mathcal
F}(x)$ comes directly from a fit to the results of the simulations given
by Shull \& van Steenberg (1985), replacing the significantly less
accurate form that is used in Chen \& Kamionkowski (2004), and in MFP06.

%{\tt AF Try to recast if possible these eqs in the same form of MFP\\
%MM A noi sembra il modo piu' sintetico di scrivere l'equazione ed e' corretto nei cfr di MFP. Infatti in MFP usavamo $\epsilon{}_{DM}$ che corrisponde a $\epsilon(z)$ nel caso in cui $f_{abs}$=1. Per scriverlo completamente alla MFP dovremmo definire anche qui $\epsilon{}_{DM}=\dot{n}_{DM}m_{DM}c^2$ e sostituire $\epsilon(z)$ con $f_{abs}\,{}\epsilon_{DM}$, che pero' e' inutile in questo articolo. Se non sei convinto, ne parliamo.}

The evolution equations in RECFAST have been modified adding the DM
energy injection terms:
\begin{eqnarray}
%\qa%begin{equation}
%\begin{array}{l}
-\delta{}\left(\frac{{\ud}x_{\rm H}}{{\ud}z}\right) & = &
\frac{\epsilon{}(z)} {E_{\rm th,H}}\,{}
\frac{1+4f_{\rm He}}{1+f_{\rm He}}\,{}
\frac{1-x_{\rm H}}{3}\,{}
\mathcal{E}\label{rec1}\\
-\delta{}\left(\frac{{\ud}x_{\rm He}}{{\ud}z}\right)& = &
\frac{\epsilon{}(z)}{E_{\rm th,He}}
\frac{1+4f_{\rm He}}{1+f_{\rm He}}\,{}
\frac{1-x_{\rm He}}{3}\,{}
\mathcal{E}\label{rec2}\\
-\delta{}\left(\frac{{\ud}T_{\rm IGM}}{{\ud}z}\right) & = &
\frac{2\,{}\epsilon{}(z)}{3\,{}k_{\rm B}}\,{}
\frac{1+4f_{\rm He}}{1+f_{\rm He}}\,{}\times\nonumber\\
&&\frac{{\mathcal F}(x_{\rm H})+f_{\rm He}\,{}{\mathcal F}(x_{\rm He})}
{1+f_{\rm He}}\,{}\mathcal{E},\label{rec3}
%\end{array}
%\end{equation}
%\nqa
\end{eqnarray} 
where $x_{\rm H}$ ($x_{\rm He}$) is the ionized fraction of
hydrogen (helium) atoms, and
${\mathcal E}\equiv{}\left[H(z)\,{}(1+z)\right]^{-1}$.  These
equations are slightly different from the ones used e.g. in Padmanabhan
\& Finkbeiner (2005) because in our case $\epsilon(z)$ is the energy
absorption rate {\it per baryon}, rather than {\it per hydrogen atom}.
%[Finally, $\epsilon{}(z)$ is
%the rate of energy absorption per baryon in the IGM, that we derived by
%integrating the equation (\ref{energy injection general}).]

We apply this formalism to two different DM candidates, i.e. sterile
neutrinos and LDM, which are expected to have the maximum impact on
reionization and heating (MFP06). In the case of sterile neutrinos only
the decay process is allowed (see Section \ref{sterile nu
subsection}). For LDM particles (Section \ref{LDM subsection}) we
discuss both the decay and the annihilation process.

We do not consider heavier DM candidates (such
as gravitinos or neutralinos) because MFP06 have already showed that,
even assuming $f_{\rm abs}=1$, their contribution to reionization and
heating is completely negligible, and there is no point in extending our
formalism in order to account for them.

%For completeness, we also considered heavier DM candidates, such as gravitinos and neutralinos (Section 3.3), although their contribution to reionization and heating is negligible.

\subsection{Sterile neutrinos}\label{sterile nu subsection}
Sterile neutrinos are one of the most popular warm DM
candidates (Colombi, Dodelson \& Widrow 1996; Sommer-Larsen \& Dolgov
2001). Their existence is predicted by the standard oscillation theory
and required by various extensions of the Minimal Standard Model, such
as the $\nu{}$MSM (Shaposhnikov 2006 and references therein). They are
massive; so they can decay following different channels. In this
paper we will consider only the so called radiative decay, \ie\ the
decay of a sterile neutrino into an active neutrino and a photon.

The mass of radiatively decaying sterile neutrinos can be constrained by
the absence of any detection of X-ray lines consistent with photons due
to sterile neutrino decays in nearby galaxy clusters (Abazajian, Fuller \& 
Tucker 2001; Abazajian 2006; Abazajian \& Koushiappas 2006; Boyarsky et al. 2006b). Recently, Watson et al. 2006 applied the same method to the X-ray emission from the Andromeda galaxy, finding an upper limit 
\begin{equation}\label{casey}
m_{\nu{}\,{}s}\,{}c^2\lesssim{}2.1 \textrm{ keV }\left(\frac{\sin{}^22\theta{}}{10^{-7}}\right)^{-0.213}, 
\end{equation}
where $\theta{}$ is the mixing angle. This limit is valid for masses $m_{\nu{}\,{}s}\,{}c^2\lesssim{}24$ keV. The most stringent constraints for masses $m_{\nu{}\,{}s}\,{}c^2>24$ keV come from the comparison between the unresolved X-ray background and the expected contribution from sterile neutrino decays (Mapelli \& Ferrara 2005; Boyarsky et al. 2006a):
\begin{equation}\label{boyarsky}
m_{\nu{}\,{}s}\,{}c^2\lesssim{}25 \textrm{ keV }\left(\frac{\Omega{}_{\rm DM}}{0.198}\right)^{-0.2}\,{}\left(\frac{\sin{}^22\theta{}}{1.55\times10^{-11}}\right)^{-0.2}. 
\end{equation}

For masses lower than $\sim{}3.5$ keV the main constraints arise from the positivity of the lepton number.

On the other hand, the study of matter power spectrum fluctuations provides a
conservative lower limit of $m_{\nu{}\,{}s}\,{}c^2\gtrsim{}2$ keV
(Viel et al. 2005), even if more recent estimates significantly increase
this lower limit ($m_{\nu{}\,{}s}\,{}c^2\gtrsim{}14$ keV, Seljak et
al. 2006; $m_{\nu{}\,{}s}\,{}c^2\gtrsim{}10$ keV, Viel et
al. 2006). It is worth noting that these lower limits are independent of the mixing angle.

According to the X-ray observational constraints discussed above, the
minimum possible lifetime for sterile neutrino radiative decays is (Mapelli \& Ferrara 2005; MFP06):
\begin{equation}\label{lifetime1}
\tau{}_{\rm DM} = 2.23\times{}10^{27}\textrm{ s }
\left(\frac{m_{\nu\,{}s}\,{}c^2}{\textrm{10 keV}}\right)^{-5}\,{}
\left(\frac{6.6\times{}10^{-11}}{\sin{}^2 2\theta{}}\right),
\end{equation}
if $3.5\lesssim{}m_{\nu\,{}s}\,{}c^2/{\rm keV}\lesssim{}24$, and
\begin{equation}\label{lifetime2}
\tau{}_{\rm DM} = 9.67\times{}10^{25}\textrm{ s }
\left(\frac{m_{\nu\,{}s}\,{}c^2}{\textrm{25 keV}}\right)^{-5}\,{}
\left(\frac{1.55\times{}10^{-11}}{\sin{}^2 2\theta{}}\right),
\end{equation}
if $m_{\nu\,{}s}\,{}c^2/{\rm keV}\gtrsim{}24$.

The current number density of sterile neutrinos $N_{s,0}$ is proportional to
the current number density of active neutrinos [$N_a=3\,{}N_{\rm b}(0)/(11\,{}\eta{})$,
where $\eta{}$ is the baryon-to-photon density]. If we assume also that
sterile neutrinos account for all the DM, we can write the number of
sterile neutrinos per baryon as (cfr. Mapelli \& Ferrara 2005): 
\begin{eqnarray}
%\begin{array}{l}
n_{s,0} = 5.88\times{}10^5
%0.147\textrm{ cm}^{-3}
\left(\frac{\rho_{\rm crit}}{\rm 10^{-29}\;g\,cm^{-3}}\right)\nonumber{}
\left(\frac{\Omega{}_{{\rm DM}}}{0.198}\right)\\
\times{}
%\left(\frac{n_{\rm b}}{2.5\times{}10^{-7}\textrm{cm}^{-3}}\right)
\left(\frac{m_{\nu{}\,{}s}\,{}c^2}{\textrm{8 keV}}\right)^{-1}
\left(\frac{\eta{}}{6.13\times{}10^{-10}}\right)^{-1}.
\end{eqnarray}
where $\rho{}_{{\rm crit}}\simeq 1.88 \times 10^{-29}\, h^2\;{\rm
g\,cm^{-3}}$ is the critical density of the Universe.

%where $\Omega{}_{{\rm DM}}=\Omega_{\rm M}-\Omega_{\rm b}$ is the current
%DM density in terms of the critical density of the Universe, and $h$ is
%the Hubble constant in units of $100\;{\rm km\,s^{-1}\,Mpc^{-1}}$. 

%For sterile neutrinos we then have $n_{{\rm DM},0}=n_{s,0}/n_{\rm b}$,
%and for each considered mass we have all the ingredients needed to calculate
%$\dot{n}_{DM}(z)$, defined in equation (\ref{DM density decay}).
Imposing $n_{{\rm DM},0} = n_{s,0}$ and $m_{\rm DM} =
m_{\nu{}\,{}s}$ we have all the ingredients needed to calculate
$\dot{n}_{\rm DM}(z)$ through equation (\ref{DM density decay}), for each
considered mass.

Each sterile neutrino decay produces one active neutrino and one photon,
each of them with an energy $\simeq {1\over2}m_{\rm DM} c^2$. Since the
active neutrino does not interact with the IGM, we only need to consider
the photon. Then, we use equations (\ref{n injection photons}) and
(\ref{e injection photons}) with $\zeta_1=1$ and $\zeta_2=1/2$, in order
to get the injection rate and average energy of photons, and proceed to
the integration of equation (\ref{energy injection general}).

%% From $\dot{n}_{DM}(z)$ we derive the injection rate of new product
%% particles (per baryon) $\dot{n}_{new}(z)=\zeta{}_1\,{}\dot{n}_{DM}(z)$,
%% where $\zeta{}_1=2$ in the case of sterile neutrinos, because each decay
%% produces a photon and an active neutrino. Similarly the average energy
%% of each product particle is $E_{new}=\zeta{}_2\,{}m_{DM}\,{}c^2$, where
%% $\zeta{}_2=1/2$, if we assume that half of the mass-energy of the DM
%% particle is given to the active neutrino and half to the photon. Because
%% the active neutrino does not interact with the IGM, we have to consider
%% only the emitted photon (which interacts via Compton scattering and
%% photo-ionization), in order to derive the rate of energy absorption per
%% baryon, $\epsilon{}(z)$, by solving the equation
%% (\ref{energy injection general}).

\begin{figure}
\center{{
\epsfig{figure=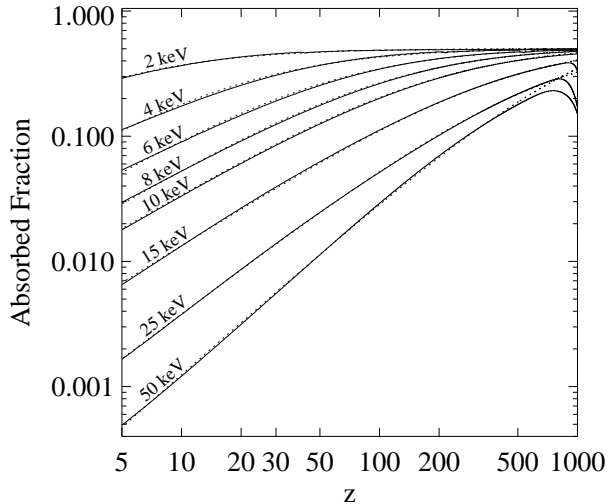,height=6.7cm}
}}
\caption{\label{fig:fig1} Absorbed fraction as a function of redshift
for sterile neutrinos of masses between 2 and 50 keV (solid lines). The
dotted lines, mostly superimposed to the solid ones, show the fitting
functions listed in Appendix B. The decrease in $f_{\rm abs}$ which can
be seen for $z\gtrsim800$ for the 15, 25 and 50 keV curves might be an
artifact caused by our choice of the redshift where the integration is
started ($z_{\rm max}=1100$), as is discussed Appendix B.}
\end{figure}

The resulting absorbed fraction, $f_{\rm abs}(z)$, is shown in
Fig. \ref{fig:fig1}. In the case of complete and immediate absorption of
the photon energy (i.e. the case studied by MFP06) $f_{\rm abs}(z)$
would be $0.5$, because half of the sterile neutrino mass-energy is
taken away by the active neutrino. It is clear that the complete
absorption approximation is pretty good at high redshift, especially for
low mass sterile neutrinos ($m_{\nu{},\,{}s}\,c^2 \lesssim{}4$ keV);
whereas at low redshift it fails by a possibly large factor. In Appendix
B, we provide analytical fits to the $f_{\rm abs}(z)$ curves shown
above, for $3\leq z\leq 1000$.
%%%%%%%%%%%%%%%%%%%%%%%%%%%%%%%%%%% FIGURE 1 %%%%%%%%%%%%%%%%%%%%%%%%%%%%%%%%%%
%%%%%%%%%%%%%%%%%%%%%%%%%%%%%%%%%%%%%%%%%%%%%%%%%%%%%%%%%%%%%%%%%%%%%%%%%%%%%%%

%% The absorbed fraction can be fit by the following equation, within an
%% error less than 4\% for $z\ge{}5$ (see Appendix B). At lower redshift
%% ($z\ge{}3$) the error is of the order of 10-20 \%.
%% \begin{equation}\label{neutrino fit}
%% f_{\rm abs}(z)=\left[0.5+0.032\,{}\left(\frac{m_{\nu{},\,{}s}}{8\textrm{ keV}}\right)^{1.5}\right] \,{} \left[\frac{z}{110\left(\frac{m_{\nu{},\,{}s}}{8\textrm{ keV}}\right)^{2.4}+z}\right]^{0.93}
%% \end{equation}

Implementing $\epsilon{}(z)$ in RECFAST (equations
\ref{rec1}-\ref{rec3}), we derive the effective influence of sterile
neutrinos on the ionization fraction and IGM temperature
(Fig.~\ref{fig:fig2}). The Thomson optical depth, $\tau{}_e$, shown in
Fig.~\ref{fig:fig2} and in the following figures, has been calculated by
integrating the well-known
formula:
\begin{equation}\label{Thomson}
\tau{}_e=
\int_{z_1}^{z_2}dz\frac{dt}{dz}\,{}c\,{}\sigma{}_T\,{}\, x\, N_{\rm b}(z),
\end{equation}
where we take  $z_2$=1000, i.e. the low-redshift boundary of the last scattering surface, and $z_1=5$, that is approximately the lowest redshift at which our fits of the absorbed fraction are valid (mostly because of our underlying assumption of a largely neutral IGM). 
%The difference with respect to integrate $\tau{}_e$ down to $z_1=0$ is less than a factor 1.3 (depending on the DM candidate).
%%%%%%%%%%%%%%%%%%%%%%%%%%%%%%%%%%% FIGURE 2 %%%%%%%%%%%%%%%%%%%%%%%%%%%%%%%%%%
\begin{figure}
\center{{
\epsfig{figure=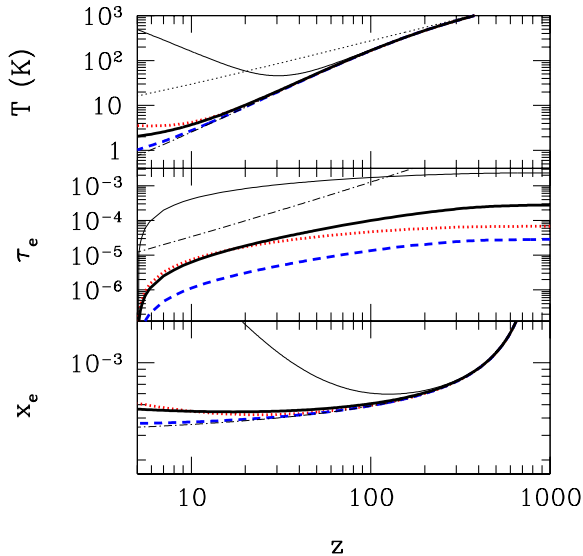,height=8cm}
}}
\caption{\label{fig:fig2}
Ionized fraction (bottom panel), Thomson optical depth (central panel)
and IGM temperature (top panel) as a function of redshift due to sterile
neutrinos. The thick lines are obtained taking into account the
effective absorbed fraction (Fig.~\ref{fig:fig1}) for sterile neutrinos
of masses 4 (thick dotted line), 15 (dashed), and 25 keV (solid). The thin
solid line shows the contribution of sterile neutrinos of mass 25~keV, if
we assume an absorbed fraction $f_{\rm abs}=0.5$.  The thin dot-dashed
line represents, from bottom to top, the relic fraction of free
electrons, their contribution to Thomson optical depth and the IGM
temperature without particle decays.  In the top panel, the thin dotted
line represents the CMB temperature.} 
\end{figure}
%%%%%%%%%%%%%%%%%%%%%%%%%%%%%%%%%%%%%%%%%%%%%%%%%%%%%%%%%%%%%%%%%%%%%%%%%%%%%%%

In Fig. \ref{fig:fig2}, the solid thin line represents the effect of
25-keV neutrinos if we assume complete absorption by the IGM ($f_{\rm
abs}=0.5$ at every $z$), whereas the solid thick line was calculated
using the derived absorbed fraction $f_{\rm abs}$.  As could be expected
from Fig.~\ref{fig:fig1}, the difference between the complete and the
effective absorption cases (i.e. the thin and the thick line) increases
as the redshift decreases. For instance, the ionization fraction in the
total absorption case (thin line) is higher by a factor $\sim{}$4 at
$z=20$, which becomes a factor $\sim{}24$ at $z=5$.  The IGM temperature
in the case of effective absorption $\epsilon{}(z)$ (thick line) is
reduced by a significant factor ($\sim{}235$ at $z=5$) with respect to
the total absorption case.  The Thomson optical depth is quite
negligible, always remaining $<10^{-3}$.

In conclusion, accounting for the effective energy absorption
significantly reduces the effect of sterile neutrino decays on
reionization and heating, when compared to the case of total
absorption (see MFP06).

The other two thick lines reported in Fig.~\ref{fig:fig2} represent the effects of sterile neutrinos of 4 (dotted line) and 15 keV (dashed), using the derived absorbed fraction $f_{\rm abs}$. One can be surprised by the fact that 4-keV sterile neutrinos have a higher impact on ionization and heating than more massive neutrinos. This result comes from two different factors. 

First of all, it depends on the fact that less massive sterile neutrinos
have higher $f_{\rm abs}$, especially at low redshift. As one can see
from the behaviour of $x_e$ and $\tau{}_e$, at high redshift 25-keV
sterile neutrinos give a stronger contribution to ionization than 4-keV
sterile neutrinos. It is only at $z\lesssim{}20$ that this tendency is reversed.
%inverted.

The second reason is not 'physical', but it depends on the state of the art of observations. In fact, we adopted for each sterile neutrino mass the shortest lifetime consistent with observations. Present-day observational constraints happen to be much stronger for a 15-keV (Watson \etal 2006) than for a 25-keV sterile neutrino (Boyarsky \etal 2006a), independently of the intrinsic properties of the decay process.

%, even if the effective absorbed fraction ($f_{\rm
%abs}$) is quite close to the maximum allowed value (0.5).

\subsection{Light dark matter}\label{LDM subsection}
We define as light dark matter (LDM) particles all the DM candidates
whose mass is between 1 and 100 MeV.
%(such as the axinos).
Such particles
have recently become of interest, because they provide a viable
explanation for the detected 511-keV excess from the Galactic centre (Kn\"odlseder et al. 2005). If they are source of the
511-keV excess, then their maximum allowed mass $m_{\rm LDM}$ should be
20 MeV, not to overproduce detectable gamma rays via internal
Bremsstrahlung (Beacom, Bell \& Bertone 2004). If we consider also the
production of gamma rays for inflight annihilations of the positrons,
this upper limit might become $\sim{}$3 MeV (Beacom \& Y\"uksel 2006).

In principle, LDM can both decay and annihilate, producing
photons, neutrinos and pairs. We will treat both LDM decays and
annihilations, making the assumption that the only decay/annihilation
products are pairs. This represents quite an upper limit, because
neutrinos do not interact with the IGM and MeV photons have a low
probability to be significantly absorbed (see Fig.~\ref{fig:fig3}, and
also the discussion in Chen \& Kamionkowski 2004).

As we did for sterile neutrinos, we assume that LDM particles compose
the entire DM.
%So the current density of LDM particles is
So the current number of LDM particles per baryon is 
\begin{eqnarray}\label{LDM density}
n_{{\rm LDM},0} = 4.46\times{}10^3
\left(\frac{\rho{}_{{\rm crit}}}{10^{-29}\textrm{g cm}^{-3}}\right)\,{}
\left(\frac{\Omega_{{\rm DM}}}{0.198}\right)\nonumber{}\\
\times{}\left(\frac{m_{{\rm LDM}}\,{}c^2}{1\,{}\textrm{MeV}}\right)^{-1}\,{}
\left(\frac{N_{\rm b}(0)}{2.5\times{}10^{-7}\textrm{ cm}^{-3}}\right)^{-1},
\end{eqnarray}
where $m_{{\rm LDM}}$ is the mass of a LDM particle.

\subsubsection{Decays}
The LDM lifetime can be derived by assuming that LDM decays produce the
detected 511-keV emission from the Galactic centre (Hooper \& Wang
2004):
\begin{equation}\label{LDM life}
\tau_{\rm DM}\sim{}
4\times10^{26}\textrm{ s}\,{}
\left(\frac{m_{\rm LDM}\,c^2}{\textrm{MeV}}\right)^{-1}.
\end{equation}
From equations (\ref{LDM density}) and (\ref{LDM life}) we can derive
$\dot{n}_{\rm DM}(z)$, defined in equation (\ref{DM density decay}).

We assume that LDM decays produce pairs.
%These pairs will
%interact with the IGM by inverse Compton scatter, collisional
%ionizations and positron annihilation.
So, the parameters needed in equations (\ref{n injection electrons}) and
(\ref{e injection electrons}) are $\zeta_1=1$ (each decay produces a
single electron and a single positron) and $\zeta_2=1/2$ (both the
electron and the positron receive approximately half of the available
energy).

\begin{figure}
\center{{
\epsfig{figure=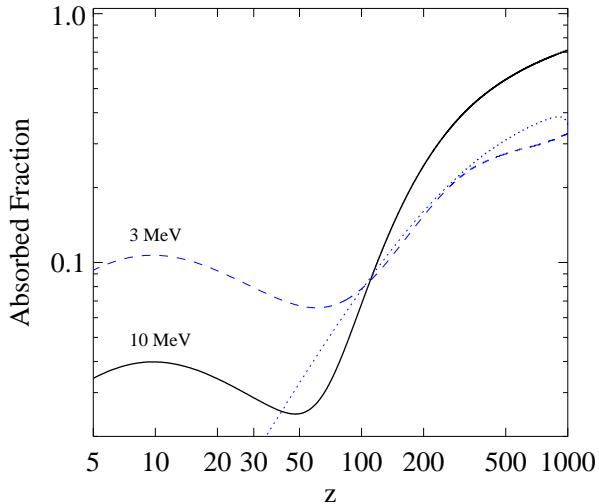,height=6.7cm}
}}
\caption{\label{fig:fig3}
Absorbed fraction as a function of redshift for LDM  particles of mass 3
(dashed line) and 10 MeV (solid line) decaying into pairs, and for LDM
particles of mass 3 MeV (dotted line) decaying into photons.  
} 
\end{figure}

Having defined these values, we then found the rate of energy absorption
per baryon, $\epsilon{}(z)$, by integrating equation (\ref{energy
injection general}). The corresponding energy absorption fraction
$f_{\rm abs}$ is shown in Fig.~\ref{fig:fig3}, where the cases of $m_{\rm LDM}\, c^2
= 3$ and 10 MeV are shown (dashed and solid line, respectively).

In the case of pair production, the assumption of immediate and complete
energy absorption corresponds to $f_{\rm abs}=1$ at every redshift,
because both the electron and the positron energy can be absorbed. The
effective value $f_{\rm abs}$ is always significantly less than 1.
%%%%%%%%%%%%%%%%%%%%%%%%%%%%%%%%%%% FIGURE 3 %%%%%%%%%%%%%%%%%%%%%%%%%%%%%%%%%%
%%%%%%%%%%%%%%%%%%%%%%%%%%%%%%%%%%%%%%%%%%%%%%%%%%%%%%%%%%%%%%%%%%%%%%%%%%%%%%%
%%%%%%%%%%%%%%%%%%%%%%%%%%%%%%%%%%% FIGURE 4 %%%%%%%%%%%%%%%%%%%%%%%%%%%%%%%%%%
\begin{figure}
\center{{
\epsfig{figure=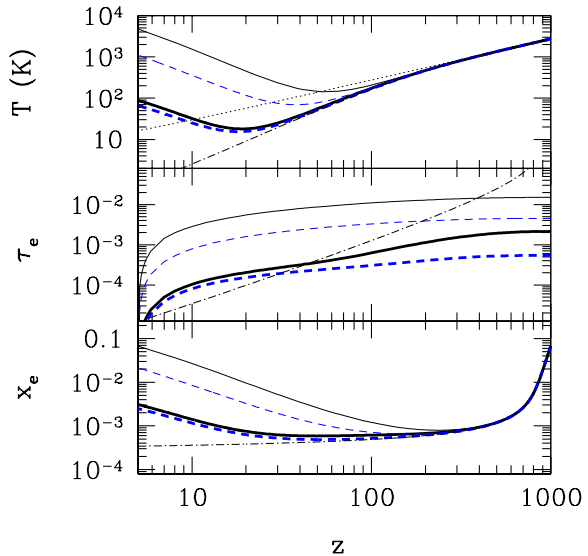,height=8cm}
}}
\caption{\label{fig:fig4}
Ionized fraction (bottom panel), Thomson optical depth (central panel)
and IGM temperature (upper panel) as a function of redshift due to
LDM decays. The thick lines are obtained using the
effective absorbed fraction (Fig.~\ref{fig:fig3}) for decaying LDM of
masses 3 (thick dashed line) and 10 MeV (solid). The thin solid (dashed)
line shows the contribution of decaying LDM of mass 10 (3) MeV, if we
assume an absorbed fraction of 1.
 The thin dotted line in the top panel and the thin dot-dashed line in all the panels are the same as in Fig.~\ref{fig:fig2}.
} 
\end{figure}
%%%%%%%%%%%%%%%%%%%%%%%%%%%%%%%%%%%%%%%%%%%%%%%%%%%%%%%%%%%%%%%%%%%%%%%%%%%%%%%

At high redshift ($z\gtrsim{}100-200$), the absorbed fraction is
relatively high (0.3-0.7, depending on the particle mass), and it is dominated by the inverse Compton scattering onto CMB photons and by the positron annihilation. %In the case of $m_{\rm LDM}\, c^2 = 10\;{\rm MeV}$, the dominant contribution comes from the inverse Compton scattering of CMB photons; in the case of $m_{\rm LDM}\, c^2 =3\;{\rm MeV}$ the positron annihilation contribution is quite significant, too.

In fact, positron annihilations contribute to $f_{\rm
abs}$ only at high redshift, because both the annihilation rate (see
equation \ref{annihilation total loss}) and the probability of
absorption of the photons they produce (equation \ref{tau e+}) scale as
positive powers of the baryon density.

Furthermore, CMB photons at high redshift are sufficiently energetic to
be scattered up to the ionization threshold, because
$\langle{}E_\gamma{}\rangle{}\sim{}30\textrm{ eV}\,{}(E/5\textrm{
MeV})^2\,{}[(1+z)/1001]$ (where $\langle{}E_\gamma{}\rangle{}$ is the
average energy of the photon after inverse Compton scattering and $E$ is
the energy of the electron/positron).  At lower redshift, the starting
energies of CMB photons is lower, and the energy boost due to the
inverse Compton is not sufficient to turn them into ionizing photons, so
the absorbed fraction drops significantly.

At low redshift ($z\lesssim{}50$), the absorbed fraction $f_{\rm abs}$
stabilizes, because collisional ionizations become dominant. However,
$f_{\rm abs}(z<50)$ is always $\sim0.1$ for 3-MeV particles, or $\sim
0.03-0.04$ for 10-MeV particles.  We derived a fit for the absorbed
fraction, reported in the Appendix B.
%\qa
%0.63*(1.+z2)^(-0.55)*exp(-(5.3/(1.+z2))^1.4) +
%  + 0.058*(1.+z2)^0.30* exp(-(141./(1.+z2))^1.4) -
%  - 0.265*exp(-(1000./(1.+z2))^1.4)
%f_{\rm abs}=0.23\,{}(1+z)^{-0.55}\,{}\exp{\left[-\left(\frac{5.1}{1+z}\right)^{1.4}\right]}\,{} +\,{} 0.222\,{}(1+z)^{0.18}\,{} \exp{\left[-\left(\frac{180}{1+z}\right)^{1.36}\right]}\textrm{ if }m_{LDM}=10\textrm{ MeV}
%\nqa

For completeness, Fig. \ref{fig:fig3} also shows the case of LDM decays
producing an active neutrino and a photon (dotted line) for 3-MeV LDM
particles. 
In this case, at high redshift ($z\gtrsim 100$), the absorbed
fraction for the LDM radiative decay is quite similar to that for the decay
into a pair (especially if $m_{\rm LDM}\lesssim{}3$ MeV). 
%In this redshift range, for higher LDM masses, pairs transfer significantly more energy to the IGM than photons; the opposite is true for lower values of $m_{\rm LDM}$. 
However, at redshifts
$z\lesssim 100$ $f_{\rm abs}$ drops to much lower values for photons
than for pairs, regardless of $m_{\rm LDM}$.

%% much lower
%% than for pairs. In particular, the absorbed fraction for photons is
%% comparable with that for pairs only at high redshift ($z\gtrsim{}100$),
%% where the Compton scattering is important. At lower redshift the
%% absorbed fraction for photons drops, because such high energy photons
%% are highly inefficient in photo-ionizing the IGM. The absorbed fraction
%% for photons produced by LDM particles with mass higher than 3 MeV is
%% even lower.

In Fig.~\ref{fig:fig4} we show the effects of LDM decays on reionization
and heating, both considering the energy absorption rate $\epsilon{}(z)$
 (thick lines) and the upper limit of complete
absorption (thin lines, $f_{\rm abs}=1$). For LDM decays the difference
between the two cases is important, even if less than for 25-keV sterile
neutrinos. For example, if $m_{\rm LDM}\, c^2$=10 MeV, the ionization
fraction (the IGM temperature) at $z\sim{}5$ is a factor $\sim{}23$
($\sim{}57$) higher in the case of total absorption than if we consider
our estimate of $f_{\rm abs}$. The Thomson optical depth is reduced by a
factor $\sim{}7$, and it is only $\tau{}_e\lesssim{}2.1\times{}10^{-3}$
(instead of $\tau{}_e\lesssim{}1.5\times{}10^{-2}$).

\subsubsection{Annihilations}
We now consider the case of LDM particles of mass $m_{\rm LDM}\,c^2=$1,
3 and 10 MeV, assuming that they annihilate and produce
electron-positron pairs. Recently Zhang et al. (2006) found that, in
order to be consistent with the 1-yr WMAP results at the $1\,\sigma$
level, the thermally averaged annihilation cross-section for LDM
annihilations must be\footnote{In eq. (\ref{sigma ann LDM}) the upper
limit actually given by Zhang et al. 2006 was multiplied by a factor of
2 in order to account for differences between the two treatments.}
%\begin{equation}\label{sigma ann LDM}
%\langle{}\sigma{}v\rangle{}=10^{-29}\textrm{ cm}^{3}\textrm{ s}^{-1}\,{}f_{\rm abs}^{-1}\,{}F_{26}\,{}\left(\frac{m_{LDM}\,{}c^2}{\textrm{MeV}}\right),
%\end{equation}
%where $F_{26}$ is the so-called annihilation intensity parameter, and
%$F_{26}\leq 1$ is required in order to be consistent with 1-yr WMAP
%within 1-$\sigma{}$. The absorbed fraction $f_{\rm abs}$ depends on the
\begin{equation}\label{sigma ann LDM}
\langle{}\sigma{}v\rangle{} \leq 2.2 \times 10^{-29}
\textrm{ cm}^{3}\textrm{ s}^{-1}\,{}f_{\rm abs}^{-1}\,{}
\left(\frac{m_{\rm LDM}\,{}c^2}{\textrm{MeV}}\right),
\end{equation}
The absorbed fraction $f_{\rm abs}$ depends on the redshift. However, as
a conservative approximation, we take the value of $f_{\rm abs}$ in
equation (\ref{sigma ann LDM}) to be the maximum value $f_{\rm abs,max}$
that we derive with our method (see Fig.~{\ref{fig:fig5}).  In
particular, $f_{\rm abs,max}\simeq0.5$ for 1 and 3-MeV LDM particles,
and $f_{\rm abs,max}\simeq0.9$ for 10-MeV particles; these values (and
the whole function $f_{\rm abs}$) are actually independent from the
value we adopt for $\langle\sigma v\rangle$.

We chose to use values of $\langle\sigma v\rangle$ which are close to
the upper limit given by the above formula, \ie 4, 12, and 24
$\times 10^{-29}\;{\rm cm^3\,s^{-1}}$ for $m_{\rm LDM}\, c^2=$ 1, 3, and
10 MeV, respectively. Such values of $\langle\sigma v\rangle$ are quite 
close to those ($\langle\sigma v\rangle$=0.3, 2.7 and 30 $\times 10^{-29}\;{\rm
cm^3\,s^{-1}}$, for the same masses) which have been inferred by
Ascasibar et al. (2006) in order to reproduce the 511-keV excess from
the Galactic centre.

From equations (\ref{sigma ann LDM}) and (\ref{LDM density}) we then
derive the rate of change of the number of DM particles per baryon
through equation (\ref{DM density annih}). In this case, the parameters
for the injection equations (\ref{n injection electrons}) and (\ref{e
injection electrons}) are $\zeta_1=1/2$, $\zeta_2=1$, because an
electron and a positron are produced for every annihilation (which obviously
involves two annihilating particles).

%%%%%%%%%%%%%%%%%%%%%%%%%%%%%%%%%%% FIGURE 5 %%%%%%%%%%%%%%%%%%%%%%%%%%%%%%%%%%
\begin{figure}
\center{{
\epsfig{figure=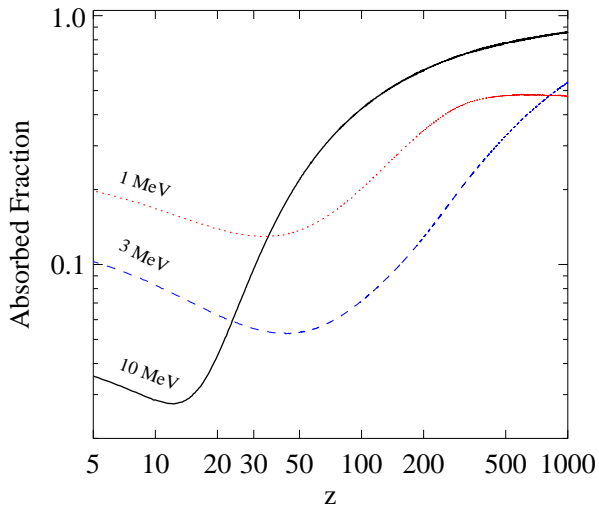,height=6.7cm}
}}
\caption{\label{fig:fig5}
Absorbed fraction as a function of redshift for annihilating LDM of mass
1 (dotted line), 3 (dashed line) and 10 MeV (solid line).  
} 
\end{figure}
%%%%%%%%%%%%%%%%%%%%%%%%%%%%%%%%%%%%%%%%%%%%%%%%%%%%%%%%%%%%%%%%%%%%%%%%%%%%%%%

The integration of equation (\ref{energy injection general}) with these
parameters provides us with the energy absorption rate per baryon,
$\epsilon{}(z)$, and the corresponding absorbed fraction $f_{\rm abs}$,
which is shown in Fig.~\ref{fig:fig5}. The behaviour of $f_{\rm abs}(z)$
is quite similar to the case of LDM decays. At high redshift
($z\gtrsim{}100$) $f_{\rm abs}(z)$ is close to the complete and
immediate absorption value ($f_{\rm abs}=1$) and it is dominated by
inverse Compton scattering and positron annihilations. At low redshift
collisional ionizations alone contribute to $f_{\rm abs}$, which suffers
a large drop between the two regimes.
%A part from that common general trend, there are some slight differences
%between LDM decays and annihilations. First of all, the energy injected
%by LDM annihilations is absorbed more efficiently at very high redshift
%with respect to LDM decays. This is mainly due to two factors. Firstly,

%The impact of positron annihilations on the absorbed fraction is more important than for LDM decays,
We note that the contribution of positron annihilations 
 to the absorbed fraction is particularly crucial for
low mass LDM particles. In fact, for the case $m_{\rm LDM}\, c^2={}$1
MeV, inverse Compton scattering is not able to produce ionizing photons,
even at $z\sim{}1000$. For this reason, in absence of positron
annihilation, the absorbed fraction for 1-MeV LDM particles (dotted line
in Fig.~\ref{fig:fig5}) would depend only on collisional ionization, and
its plot would essentially be a straight line from $f_{\rm abs}(5)\sim 0.2$ to $f_{\rm abs}(1000)\sim 0.08$.

On the contrary, for the highest mass we consider, $m_{\rm LDM}\,
c^2=10$ MeV, the high redshift bump is essentially due to inverse
Compton energy loss.
%%%%%%%%%%%%%%%%%%%%%%%%%%%%%%%%%%% FIGURE 6 %%%%%%%%%%%%%%%%%%%%%%%%%%%%%%%%%%
\begin{figure}
\center{{
\epsfig{figure=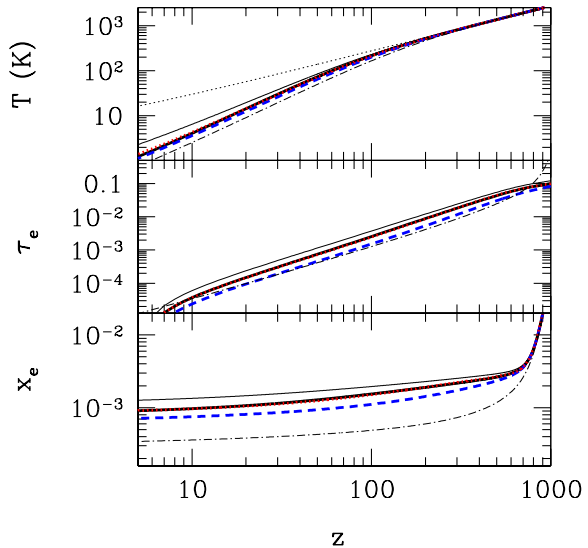,height=8cm}
}}
\caption{\label{fig:fig6}
Ionized fraction (bottom panel), Thomson optical depth (central panel) and IGM temperature (upper panel) as a function of redshift due to LDM annihilations. The thick lines are obtained taking into account the effective absorbed fraction (Fig.~\ref{fig:fig5}) for annihilating LDM of masses 1 (thick dotted line), 3 (dashed) and 10 MeV (solid). The thin solid line shows the contribution of annihilating LDM of mass 10 MeV, if we assume an absorbed fraction=1.
 The thin dotted line in the top panel and the thin dot-dashed line in all the panels are the same as in Fig.~\ref{fig:fig2}.
% The thin dot-dashed line is the same as Fig.~\ref{fig:fig2}.
} 
\end{figure}
%%%%%%%%%%%%%%%%%%%%%%%%%%%%%%%%%%%%%%%%%%%%%%%%%%%%%%%%%%%%%%%%%%%%%%%%%%%%%%%

The impact of LDM annihilations on reionization and heating
(Fig.~\ref{fig:fig6}) is quite different from the case of LDM decays. In
fact, LDM annihilations start to contribute both to reionization and
heating already at very high redshift ($z\sim{}800$); but their role
remains negligible at low redshift. In particular, the ionization
fraction becomes $\sim{}10^{-3}$ and the IGM temperature at
$z\sim{}10$ is much lower than 10 K. This mainly is due to the fact that
the annihilation rate depends on the square of the baryon density (see
equation \ref{DM density annih}).

The Thomson optical depth reported in the central panel of
Fig.~\ref{fig:fig6} is quite high: $\tau{}_e=0.08-0.10$ for all the
considered LDM particles, \ie close to the best fit
($\tau{}_e=0.09$) of the 3-yr WMAP data. Nevertheless, the effects of
LDM annihilations can be very hardly detected by WMAP. The reason is that,
differently from the relatively rapid and large variation of the
electron fraction occurring in standard reionization scenarios, the
$x_e$ evolution produced by LDM annihilations tracks very closely,
albeit at a slightly higher level, and for a long time the relic
abundance one. This behavior dilutes the effects of these extra
electrons, making their imprint on the CMB spectrum very tiny, as can be
appreciated from Fig. \ref{fig:fig7}.

%%%%%%%%%%%%%%%%%%%%%%%%%%%%%%%%%%% FIGURE 7 %%%%%%%%%%%%%%%%%%%%%%%%%%%%%%%%%%
\begin{figure}
\center{{
\epsfig{figure=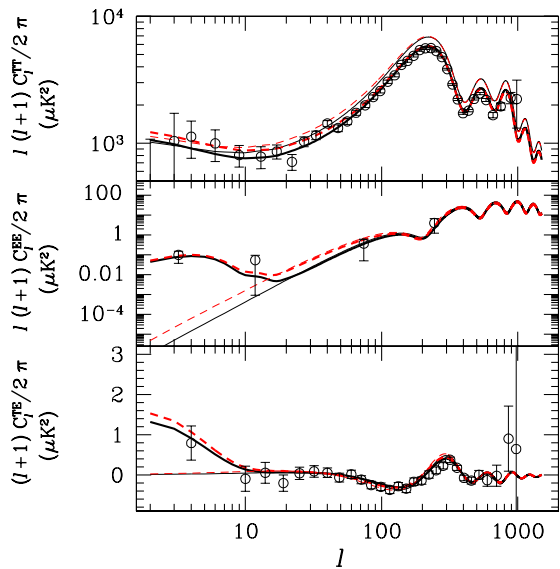,height=8cm}
}}
\caption{\label{fig:fig7} Temperature-temperature (top panel),
polarization-polarization (central panel) and temperature-polarization
(bottom panel) spectra. Thick lines indicate the CMB spectrum derived
assuming Thomson optical depth $\tau{}_e=0.09$ and a sudden reionization
model (consistent with the 3-yr WMAP data); thin lines indicate the CMB
spectrum derived assuming $\tau{}_e=0$. Dashed (solid) lines indicate
the CMB spectrum obtained (without) taking into account the
annihilations of 1-MeV LDM particles. Open circles in all the panels
indicate the 3-yr WMAP data (Hinshaw et al. 2006; Page et
al. 2006; Spergel et al. 2006).}
\end{figure}
%%%%%%%%%%%%%%%%%%%%%%%%%%%%%%%%%%%%%%%%%%%%%%%%%%%%%%%%%%%%%%%%%%%%%%%%%%%%%%%

However, the Thomson optical depth produced by LDM annihilations should
influence the CMB spectra at quite high multipoles, as implied by the
results of Zhang et al. (2006). In Fig.~\ref{fig:fig7} we show how
annihilating LDM particles of mass 1-MeV modify the
temperature-temperature (TT), polarization-polarization (EE) and
temperature-polarization (TE) CMB spectra. The spectra have been
simulated by running the public code CMBFAST (Seljak \& Zaldarriaga
1996; Seljak et al. 2003), modified in order to account for the effects
of DM decays/annihilations. The main effects of LDM annihilations are a
certain damping in the TT peaks, a sensible variation of the EE spectra
for $l\lesssim{}100$ and some negligible distortions in the TE
spectra. However, the simulated spectra agree within 1-$\sigma{}$ with
the 3-yr WMAP results.  Our plot is obtained assuming that all the
cosmological parameters have the best fit value indicated by the 3-yr
WMAP data. Leaving the cosmological parameters free to change, it should
be possible to get an even better agreement between the WMAP data and
the simulated CMB spectra derived accounting for LDM annihilations
(Zhang et al. 2006).

\section{Conclusions}
We have modelled the absorption rate of the energy released in the
IGM by DM decays/annihilations producing photons,
electron-positron pairs and neutrinos. Our model suitably describes the
energy deposition of a wide class of DM candidates, such as sterile
neutrinos and LDM.  Useful fits to the absorbed energy fraction as a
function of redshift for the various particles are given in Appendix B.

In the case of radiatively decaying sterile neutrinos (mass 2-50 keV), at
$z > 200$ a fraction $f_{\rm abs}\simeq{}0.5$ of the particle energy is
transferred to the IGM, predominantly via ionizations; at lower
redshifts $f_{\rm abs}$ decreases rapidly to values of 0.0005-0.3
depending on the neutrino mass.  LDM particles can decay
or annihilate. In both cases $f_{\rm abs}\approx 1$ at high ($> 300$)
redshift, due to positron annihilation and inverse Compton scattering,
and it drops to values around 0.1 below $z=100$.

Our determination of $f_{\rm abs}$ has a dramatic impact on the results of
previous studies (which adopted naive assumptions for this parameter)
concerned with the IGM heating by DM.

To illustrate this point, we have re-calculated the IGM thermal and
ionization history induced by either sterile neutrinos or LDM particles,
using the previous findings. We find that sterile neutrino (LDM) decays
are able to increase the IGM temperature by $z=5$ at most up to $4$~K
($100$~K). Both these values are 50-200 times lower than the estimates
based on the assumption of complete energy transfer to the gas.  In
addition, significant departures from the adiabatic temperature
evolution induced by the Hubble expansion occur only below $z \approx
30$, at an epoch when heating and
ionization by conventional sources (stars or accretion-powered objects)
are likely to swamp the DM signal.

LDM annihilations instead produce a
very extended ($5 < z < 800$) electron fraction plateau, at a level of
5-10 times the relic one. The main effect of these extra electrons is to
extend the cosmic time interval during which the IGM kinetic temperature
is coupled to the CMB one down to $z\approx 100$. Although the electron
scattering optical depth in this case is large (0.08-0.10), its effects
on the CMB temperature/polarization spectra are hardly appreciable.

%% {\bf ER: questo lo toglierei del tutto visti i commenti del referee (che
%%   dice che quelli che chiamiamo lower limits non sono lower limits..)

%% It is also remarkable that in the case of LDM annihilations the upper
%% limits on $\langle\sigma v\rangle$ which can be obtained by combining
%% the Zhang et al. (2006) results with our determination of $f_{\rm abs}$
%% are quite close to the results of Ascasibar et al. (2006), which instead
%% represent a lower limit for the annihilation cross section. In
%% particular, in the case of a 10 MeV annihilating LDM particle the upper
%% limit obtained from the Zhang \etal (2006) formula is {\it lower} than
%% the Ascasibar et al. (2006) limit, although the discrepancy is only at
%% the $1-\sigma$ level.}

The detailed computation of the $f_{\rm abs}$ presented in this paper and
summarized by the fits given in the Appendix B, might be useful for a
large number of future applications in which the cosmological role of
the DM is investigated.  Among these are the effects of DM
decays/annihilations on the 21 cm emission (Shchekinov \& Vasiliev 2006)
and on the structure formation history (Shchekinov \& Vasiliev 2004;
Biermann \& Kusenko 2006; Ripamonti, Mapelli \& Ferrara 2006).

%gravitinos/neutralinos

%%%%%%%%%%%%%%%%%%%%%%%%%%%%%%%%%%% TABLE 1 %%%%%%%%%%%%%%%%%%%%%%%%%%%%%%%%%%%
%\begin{table}
%\begin{center}
%\caption{Initial parameters for the Milky Way model.
%}
%\begin{tabular}{ll}
%\hline
%\hline
%\vspace{0.1cm}
%$c$ & 12\\
%$V_{200}$ & 160 km s$^{-1}$\\
%\hline
%\end{tabular}
%\end{center}
%\label{tab_1}
%\end{table}
%%%%%%%%%%%%%%%%%%%%%%%%%%%%%%%%%%%%%%%%%%%%%%%%%%%%%%%%%%%%%%%%%%%%%%%%%%%%%%%

\section*{Acknowledgements}
We thank S. Zaroubi, A. Kusenko, P. Biermann, J. Stasielak, C. Watson
and X. Chen for useful discussions, and for detailed explanations about
their results. We also thank the referee, Y. Ascasibar, for a careful
reading of the manuscript and useful comments. ER (MM) thanks SISSA/ISAS
(the Kapteyn Institute) for the hospitality during the preparation of
this paper. ER gratefully acknowledges support from the Netherlands
Organization for Scientific Research (NWO) under project number 436016.

%Formato per le figure
%\begin{figure}
%\center{{
%\epsfig{figure=name.ps,height=8cm}
%}}
%\caption{\label{name} caption}
%\end{figure}

\onecolumn

\appendix

\section{Internal Bremsstrahlung}
In Section 2.2 we have considered the energy losses of electrons and
positrons produced by DM decays or annihilations. However, because of
the so called {\it internal Bremsstrahlung} process, it is possible that
an annihilation results in the production of an electron-positron pair,
plus a photon (Beacom et al. 2005). As the photon carries away a
fraction of the energy released by the annihilation, this mechanism
could influence $f_{\rm abs}$.

We can estimate the fraction of energy carried away by the photons
produced by internal Bremsstrahlung by using the cross-section reported
in Beacom et al. (2005)
\begin{equation}
{{d\sigma_{\rm Br}}\over{dE}} =
\sigma_{\rm tot}\, {\alpha\over\pi}\, {1\over E}\,
\left[{\ln \left({{s'}\over{m_{\rm e}^2\,c^4}}\right)-1}\right]
\left[{1+\left({{s'}\over s}\right)^2}\right]
\end{equation}
where $E$ is the energy of the photon, $\sigma_{\rm tot}$ is the total
annihilation cross-section in the three-level approximation (Peskin \&
Schroeder 1995), $\alpha$ is
the fine structure constant, $s=4\,{}(m_{\rm DM}\,c^2)^2$, and $s'=4\,m_{\rm
DM}\,c^2\, (m_{\rm DM}\,c^2-E)$.

Then, the fraction of the annihilation energy which is carried away by
the photons is
\begin{equation}
f_{\rm Br} = {1\over{2\,\sigma_{\rm tot}\,m_{\rm DM}\,c^2}}
\int_0^{m_{\rm DM}\,c^2} {{{d\sigma_{\rm Br}}\over{dE}}\, E\, dE}\ \leq
{{2\alpha}\over\pi} \ln(2 m_{\rm DM}/m_{\rm e})
\end{equation}
where we have exploited the fact that
\begin{equation}
{{d\sigma_{\rm Br}}\over{dE}}\ \leq
\sigma_{\rm tot}\,{{4\alpha}\over\pi}\,
{{\ln(2 m_{\rm DM}/m_{\rm e})}\over E}.
\end{equation}
So, even for the highest mass we consider ($m_{\rm DM}=10\;{\rm MeV}$),
$f_{\rm Br}\leq0.017$. Such a fraction is small; but, if the internal
Bremsstrahlung photons were completely absorbed, it is enough to
significantly alter the value of $f_{\rm abs}$ at low redshift, at least
for 10-MeV LDM annihilations (see Fig. 5).

However, this is not the case: the typical energy of a photon produced
by the internal Bremsstrahlung mechanism is $\sim {1\over2}m_{\rm
DM}c^2$, and equation (\ref{tau e+}) can be used in order to estimate
the optical depth encountered by such a photon for each Hubble length it
travels. For the 5-MeV photons typically produced by the annihilations
of 10-MeV LDM particles this is $\tau\simeq2.2\times10^{-4}
(1+z)^{3/2}$, so that the complete absorption scenario is realistic only
for $z\gtrsim150$, when $f_{\rm Br}\ll f_{\rm abs}$. Then, it is
justified to neglect the internal Bremsstrahlung process in estimating
$f_{\rm abs}$.

\section{Fits of the absorbed energy fraction}
It can be useful to derive a fit of the energy absorbed fraction
($f_{\rm abs}$) for each considered DM particle. We calculated fits with
errors smaller than 5\% for $5\lesssim{}z\lesssim{}1000$. In the
redshift range $3\lesssim z\lesssim 5$ the errors are larger ($\lesssim
10-20\%$); however, at such low redshift our absorbed fractions are likely
inaccurate by a larger factor, as our assumption of a mostly neutral IGM
breaks down.

%% The absorbed fraction can be fit by the following equation, within an
%% error less than 4\% for $z\ge{}5$ (see Appendix B). At lower redshift
%% ($z\ge{}3$) the error is of the order of 10-20 \%.
%% \begin{equation}\label{neutrino fit}
%% f_{\rm abs}(z)=\left[0.5+0.032\,{}\left(\frac{m_{\nu{},\,{}s}}{8\textrm{ keV}}\right)^{1.5}\right] \,{} \left[\frac{z}{110\left(\frac{m_{\nu{},\,{}s}}{8\textrm{ keV}}\right)^{2.4}+z}\right]^{0.93}
%% \end{equation}

\subsection{Sterile neutrinos}

For the case of sterile neutrino with mass $2\,{}\leq{}m_{\nu,s}\,c^2/\textrm{ keV}\leq 10$, a fit was found depending only on the redshift and the mass of
the particle.
\begin{equation}
f_{\rm abs}(z,m_{\nu,s})=
\left[0.5+0.032\,{}
\left(\frac{m_{\nu{},\,{}s}\, c^2}{8\textrm{ keV}}\right)^{1.5}\right] \,{}
\left[\frac{z}{110
\left(\frac{m_{\nu{},\,{}s}\,c^2}{8\textrm{ keV}}\right)^{2.4}+z}\right]^{0.93}.
\end{equation}

This general formula does not hold for higher masses, for which we have found
\begin{eqnarray}
f_{\rm abs}(z,{\rm 15 keV}) & = & 
0.99\,\left[0.5+0.032\,{}
\left(\frac{15\textrm{ keV}}{8\textrm{ keV}}\right)^{1.5}\right] \,{}
\left[\frac{z}{110
\left(\frac{14\textrm{ keV}}{8\textrm{ keV}}\right)^{2.4}+z}\right]^{1.0}
\nonumber \\
f_{\rm abs}(z,{\rm 25 keV}) & = &
0.89\,\left[0.5+0.032\,{}
\left(\frac{25\textrm{ keV}}{8\textrm{ keV}}\right)^{1.5}\right] \,{}
\left[\frac{z}{110
\left(\frac{17\textrm{ keV}}{8\textrm{ keV}}\right)^{2.4}+z}\right]^{1.2}
\nonumber \\
f_{\rm abs}(z,{\rm 50 keV}) & = &
1.08\,\left[0.5+0.032\,{}
\left(\frac{50\textrm{ keV}}{8\textrm{ keV}}\right)^{1.5}\right] \,{}
\left[\frac{z}{110
\left(\frac{22\textrm{ keV}}{8\textrm{ keV}}\right)^{2.4}+z}\right]^{1.4}
\end{eqnarray}

As can be seen in Fig. \ref{fig:fig1}, these last fits do not take
into account the decrease of $f_{\rm abs}$ at very high redshift, which
is caused by the fact that we start integrating equation (\ref{energy
injection general}) at redshift $1100$. In fact, if we start the
integration at $z=1500$, the high redshift discrepancies between the
fitting formulae above and the actual $f_{\rm abs}$ disappear; however,
such a procedure is uncertain because the equations we use for
describing the energy absorption might not be applicable at an epoch
before recombination.

Luckily, the energy injection from sterile neutrino decays is completely
negligible at such high redshift, and this uncertainty by a factor
$\lesssim2$ can be safely ignored.

\subsection{Light Dark Matter}
For LDM particles decaying into pairs, we derived two different fits,
for 3 and 10-MeV particles, respectively.
\begin{eqnarray}
f_{\rm abs}(z,\,{}3\textrm{ MeV}) & = &
0.49\,{}(1+z)^{-0.5}\exp{\left[-\left(\frac{5}{1+z}\right)^{1.4}\right]}+
\nonumber\\
& & \quad + \,{} 0.058\,{}(1+z)^{0.28}\,{}
\exp{\left[-\left(\frac{164}{1+z}\right)^{1.4}\right]}\,{}
-\,{} 0.265\,{}
\exp{\left[-\left(\frac{1220}{1+z}\right)^{1.4}\right]}\nonumber{}\\
f_{\rm abs}(z,\,{}10\textrm{ MeV}) & = &
0.21\,{}(1+z)^{-0.55}\,{}
\exp{\left[-\left(\frac{5.1}{1+z}\right)^{1.4}\right]}\,{}+
%\nonumber\\
%& & \quad +\,{}
0.222\,{}(1+z)^{0.18}\,{}
\exp{\left[-\left(\frac{185}{1+z}\right)^{1.36}\right]}
%\textrm{ if }m_{LDM}=10\textrm{ MeV}
\end{eqnarray}

Finally, for annihilating LDM particles,  we derived the following fits.
\begin{eqnarray}
f_{\rm abs}(z,\,{}1\textrm{ MeV}) & = &
0.32\,{}(1+z)^{-0.27}\,{}+\,{}
0.55\,{}(1+z)^{0.06}\,{}
\exp{\left[-\left(\frac{195}{1+z}\right)^{0.9}\right]}
%\nonumber\\
%& & \quad
\times\,{}\exp{\left[-\left(\frac{1+z}{1900}\right)^{1.2}\right]}
\nonumber\\
f_{\rm abs}(z,\,{}3\textrm{ MeV}) & = &
0.21\,{}(1+z)^{-0.39}\,{} + \,{}
+ 0.155\,{}(1+z)^{0.20}\,{}
\exp{\left[-\left(\frac{350}{1+z}\right)^{0.7}\right]}\,{}+\nonumber\\
& & \quad+\,{}
2.9\times{}10^{-4}\,{}(1+z)\,{}
\exp{\left(-\frac{550}{1+z}\right)}\nonumber{}\\
f_{\rm abs}(z,\,{}10\textrm{ MeV}) & = & 0.064\,{}(1+z)^{-0.34}\,{} +\,{} 0.335\,{}(1+z)^{0.14}\,{}\exp{\left[-\left(\frac{52}{1+z}\right)^{1.3}\right]}
\end{eqnarray}

\end{document}